# Magnet R&D for the Muon Collider
# European Strategy Input

L. Bottura, B. Auchmann, F. Boattini, B. Bordini, B. Caiffi, L. Cooley, S. Fabbri, S. Gourlay, S. Mariotto, T. Nakamoto, S. Prestemon, M. Statera

## Abstract

The Muon Collider, as proposed by the International Muon Collider Collaboration (IMCC), represents a groundbreaking advancement in circular collider technology. By using muons instead of protons or electrons, this collider has the potential of unprecedented discovery reach, luminosity, and compact design, significantly increasing energy efficiency, reducing environmental impact and improving sustainability. However, achieving this vision necessitates overcoming unique and extreme challenges in superconducting magnet technology. This document summarizes the state of the art, challenges, and the proposed R&D roadmap for developing the next generation of superconducting magnet systems crucial for the Muon Collider over the next ten years. The goal is to advance accelerator magnet technology beyond current limits, with a special focus on High-Temperature Superconductors (HTS) materials for high-field and high-temperature applications. This note is a concise summary of the extensive proposal [1] which we refer to for detailed referencing and as supporting material. We focus here on the technology gap to be filled by the proposed R&D, the structure and objectives of the proposed R&D, and provide a resource estimate for the next ten years.

Contact author: Luca Bottura, CERN ATS-DO, CH-1211 Geneva 23, Switzerland, Luca.Bottura@cern.ch

**Context**

The Muon Collider (MC) [2, 3] embodies a groundbreaking concept in circular colliders for high-energy physics, offering a unique pathway to achieve unprecedented energy, luminosity and efficiency (cost per parton collision), while significantly reducing footprint and environmental impact compared to conventional collider technologies. A critical aspect of its feasibility lies in the development of cutting-edge superconducting magnet systems capable of meeting the demanding requirements of muon production, acceleration, and collision. This was already recognized in the seminal US-based Muon Accelerator Program (US-MAP) study [4, 5] which produced an initial configuration that was crucial to identify the main challenges. In recent years, the International Muon Collider Collaboration (IMCC) [6], hosted at CERN, has evolved significantly this initial concept. The integrated design effort of IMCC was conducive to significant progress in both the conceptual and, in some cases, the engineering design of the magnet systems [7, 8]. The status of magnet development and other main findings can be found in references [9-23], and in the extended summary of [24] where we have reviewed performance requirements and identified the main challenges associated with the concept baseline.

**Objective**

In this paper we recall the development targets for the Muon Collider magnets and highlight the associated challenges that are the main development drivers. We then briefly review the state of the art in solenoid, dipole and quadrupole magnets built with LTS and HTS technology, and we conclude with a gap analysis to identify and motivate the priority directions for future R&D. We then detail the R&D proposal, built to foster magnet technology for a muon collider. The majority of the targets of the R&D proposed are well beyond the present state of the art and are not pursued by other R&D or studies in HEP.

**Methodology**

<u>Set development targets and identify key technological challenges</u>

The study of the machine configurations produced by the US Muon Accelerator Program (US-MAP) [4, 5], and the evolution in beam physics of the last few years fostered by the International Muon Collider Collaboration (IMCC) [6] and the EU Design Study MuCol, have led to the formulation of clear performance specifications and the identification of technical challenges for the whole muon collider complex. A concise summary of the magnet development targets is reported in Tab. I. Based on the figures reported there, we have identified the following broad categories of challenges:

- *High Magnetic Field Performance* – The Muon Collider will require solenoids producing magnetic fields up to 40T for final cooling and 20T for the target solenoids.

Dipole and quadrupole magnets in the accelerator must achieve bore fields up to 14T and gradients of 300T/m, in steady-state conditions.

- *Large Bore, Forces and Stored Energy* – Solenoid magnets in the target area must accommodate large apertures, up to 1400 mm bore, while storing 1.4 GJ in the whole system. Solenoids for the 6D cooling have a range of apertures, with upper values up to 800 mm bore, and up to 75 MJ stored energy in a single solenoid. Finally, collider magnets have a wide range of apertures that are large if compared to present state-of-the-art, from 140 mm dipoles in the arc up to 500 mm quadrupoles in the interaction regions, with stored energy of several tens of MJ per single magnet. The high stored energy and associated energy density up to 300 MJ/m$^3$ (accelerator magnets need to be built with compact winding for cost reasons, see also next bullet) makes quench protection a challenge. The high field and large dimensions are associated with exceptional electro-magnetic loads and stresses, e.g. 400 MPa transverse pressure on the superconductor, and 600 MPa in structural components.
- *Superconducting Magnets* of *Minimal Cost and Maximum Energy Efficiency* – Capital Expenditure (CAPEX) and Operating Expenditure (OPEX) are good indicators of the efficiency of the resources invested, and of the power required to run the collider complex. The Muon Collider will require compact, high engineering current density HTS coils, well above 500 A/mm$^2$ overall, to reduce material costs (CAPEX). In most cases the coil will be operated at a target temperature of 20 K, to deal efficiently with the cryogenic heat loads from muon decay (OPEX). Both current density and

Table I. Summary of magnet development targets for the Muon Collider.

| | | Target, decay and capture | 6D cooling | Final cooling | Rapid cycling synchrotron | | Collider ring | | | |
|---|---|---|---|---|---|---|---|---|---|---|
| Magnet type | (-) | Solenoid | Solenoid | Solenoid | NC Dipole | SC Dipole | Dipole | Dipole | Dipole | Quadrupole |
| SC material options | (-) | HTS | HTS/LTS[2] | HTS | N/A | HTS | Nb-Ti | Nb$_3$Sn | HTS | HTS |
| Aperture | (mm) | 1400 | 60…800[3] | 50 | 30x100 | 30x100 | 160 | 160 | 140 | 140 |
| Length | (m) | 19 | 0.08…0.3[3] | 0.5…1[4] | 5 | 2 | 4…6[4] | 4…6[4] | 4…6[4] | 3…9[4] |
| Number of magnets | (-) | 20 | 2 x 3030 | 20 | 7000[6] | 3000[6] | 1250[8] | 1250[8] | 1250[8] | 28 |
| Bore Field/Gradient | (T)/(T/m) | 20 | 2.6…17.9[3] | > 40 | ± 1.8[5] | 10 | 5 | 11 | 14 | 300 |
| Ramp-rate | (T/s) | SS | SS | SS | 3320…810[7] | SS | SS | SS | SS | SS |
| Stored energy | (MJ) | 1400 | 5…75 | 4 | 0.03 | 3.4 | 5 | 20 | 24 | 60 |
| Heat load | (W/m) | 2[1] | TBD | TBD | 1200 | 5 | 5 | 5 | 10 | 10 |
| Radiation dose | (MGy) | 80 | TBD | TBD | TBD | TBD | 30 | 30 | 30 | 30 |
| Operating temperature | (K) | 20 | 20 | 4.5 | 300 | 20 | 4.5 | 4.5 | 20 | 4.5…20 |

NOTES:
(1) Intended as linear heat load along the conductor wound in the solenoid. Total heat load in the target, decay and capture solenoid is approximately 4 kW.
(2) Superconducting material and operating temperature to be selected as a function of the system cost. Present baseline study is oriented towards HTS at 20 K.
(3) The range indicated covers the several solenoid magnet types that are required for the cooling cells. Extreme values typically do not occur at the same time.
(4) Specific optics are being studied, the length range indicated is representative.
(5) Rapid Cycled Synchrotrons require uni-polar swing, from zero to peak field. Hybrid Cycled Synchrotrons require bi-polar swing, from negative to positive peak field
(6) Considering the CERN implementation (SPS+LHC tunnels)
(7) Required ramp-rate decreases from the first to the last synchrotron in the acceleration chain
(8) Considering a collider of the final size (approximately 10 km length)

- operating temperature targets will require substantial advances in magnet technology, priming R&D on HTS solutions, as well as cryogenic cooling technology. We note here that operation at high temperature will also serve to address the risk in the helium supply chain.
- *Fast Pulsed Synchrotrons* – The Rapid- and Hybrid-Cycled Synchrotrons (RCS and HCS) in the muon acceleration sequence (three or four depending on the implementation) rely on pulsed resistive magnet systems capable to generate 1.8 T bore field with sub-ms rise time and 5 Hz repetition rate. The main issue is the management of the stored magnetic energy, several tens of MJ per synchrotron, which will need highly efficient energy recovery and powering systems to ensure sustainable operation. Though specific to the RCS and HCS, this is a second case where the R&D will need to control and minimize CAPEX and OPEX.
- *Radiation Resistance* – The collider magnets must withstand radiation doses exceeding 80 MGy and several $10^{-3}$ DPA's , which necessitates radiation-resistant materials and is of concern for the performance of superconducting materials, HTS most of all.

State of the Art and Gap Analysis

*Current State of Superconducting Magnet Technology*

Superconducting magnet technology has advanced significantly in recent decades, with notable progress in Low-Temperature Superconductors (LTS) such as Nb-Ti and $Nb_3Sn$, as well as High-Temperature Superconductors (HTS) like REBCO. As to accelerator magnets, the Large Hadron Collider (LHC) has successfully deployed NbTi-based dipoles operating at 8.3 T, while future projects such as the High Luminosity LHC (HL-LHC) are incorporating $Nb_3Sn$ accelerator magnets capable of reaching 11T. In the realm of solenoid technology, hybrid superconducting magnets combining LTS and HTS have demonstrated fields exceeding 30T, with promising early-stage results for all-HTS designs. As to pulsed accelerator magnet systems, the performance of single magnet and power units is within reach: ramped accelerator magnets with bore field of 1.8 T are commonplace, and fast pulsed system reaching field ramp rate in excess of 10 kT/s are the technical solution for fast kickers. The challenge is the integration of magnet performance, energy recovery and power control. The magnet and powering system proposed for the muon collider has similarities with the J-PARC RCS dipoles [25], which operate at 1.1 T peak field, 6.1 kA peak current, and ramp in approximately 20 ms (about 70 T/s) at 25 Hz.

*Technology Gaps*

While the achievements quoted above provide a strong foundation, key technology gaps remain for the magnets of the Muon Collider. Driven by the collider performance requirements, we are focusing on HTS technology for all superconducting magnets, considering LTS, Nb-Ti or $Nb_3Sn$, only for the collider accelerator magnets. We have reported in Fig. 1 an evaluation of the Technology Readiness Level (TRL) [26] of HTS and LTS magnet technology for the specific magnet types of the Muon Collider. The HTS magnets are at TRL 3 to 4, because they are still at the level of laboratory development and demonstration. The exception is the solenoid form the target and capture channel, which can rely on the manufacturing solutions developed for fusion applications. LTS magnets have higher readiness, TRL 6 to 7 for $Nb_3Sn$ and TRL 7 to 8 for Nb-Ti, but they fall short of the needs of a full performance Muon Collider. Resistive magnets and pulsed power converters for the accelerators also have high TRL at a single component level, TRL 6 to 8. At system level, with the demands of a full performance Muon Collider, they are significantly lower, TRL 4 to 5.

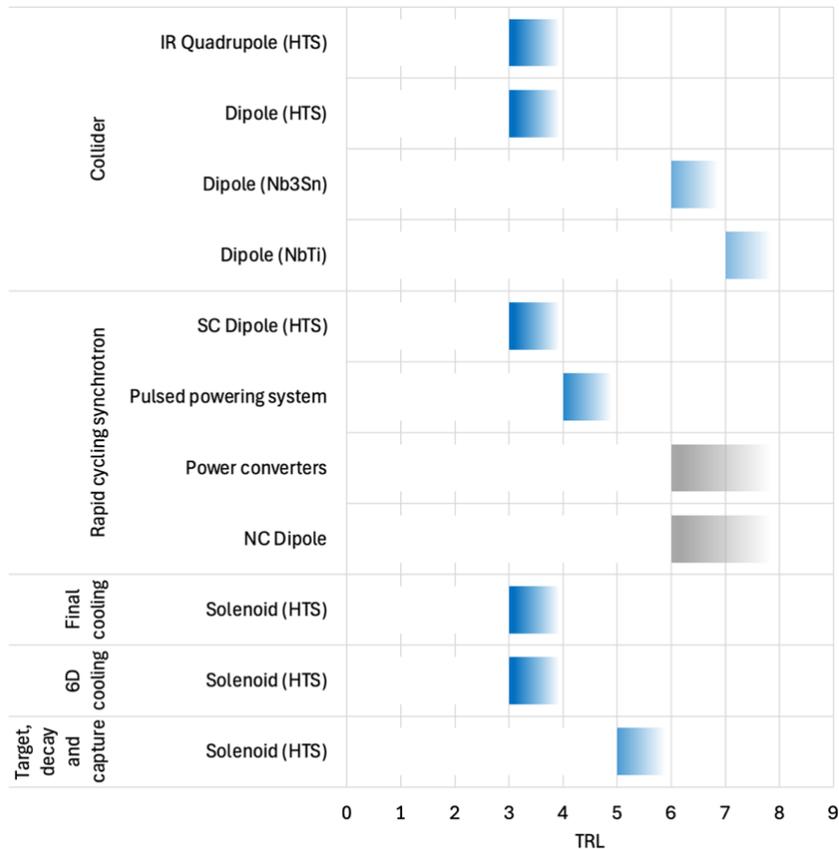

Figure 1. Summary of estimated TRL for the magnet concepts considered for the muon collider. Normal Conducting (NC) dipoles and power converters for the rapid cycling synchrotrons are quoted separately (grey area) and as a system. A secured TRL of 6 is needed for a decision for construction and is set as the objective of the R&D plan.

It is commonly accepted that a decision on technology insertion and technology transition requires a TRL 6, i.e. engineering-scale models or prototypes tested in a relevant environment. This is also the level required by the US-DOE to start construction of a new infrastructure [27]. The proposed R&D program aims at filling the gap between the present TRL and achieve TRL 6 by the end of the activities.

## Timeline and resources

The proposed R&D roadmap is structured as a development plan culminating in full-scale demonstration and prototype testing. The strategy is organized into eight lines of activities directed to the achievement of Technology Milestones (TMs) that demonstrate the required level of TRL. Each TM is associated with a specific research objective and a demonstrator, either a magnet or a system test (for the Rapid Cycling Synchrotron). Besides validating the required TRL, the demonstrators serve as focal points for the R&D of a given line of activity towards the corresponding TM, helping to streamline and rationalize the work. While activities towards each TM are distinct, several TMs are potentially interrelated by common technology, with knowledge and advancements from one line of activity feeding into others to accelerate overall progress. The program is defined over a ten-year time span with the TMs related to solenoids, LTS accelerator magnets and the pulsed resistive magnet powering achieved within this time span. HTS accelerator magnet R&D is expected to extend beyond this time frame, as there is no pressing need to deliver results within the anticipated timeline for a potential Muon Collider.

Table II. Overview of lines of activity towards the eight Technology Milestones of the proposed R&D program. Also reported the time of the main deliverable and the resources associated (M+P)

| Technology | Technology Milestone Demonstrator | Objectives and Deliverables | Key Parameters and Targets | Time | Resources |
|---|---|---|---|---|---|
| Solenoid for target, decay and capture channel | Target solenoid model coil (20@20) | Develop conductor, winding and magnet technology suitable for a target solenoid, generating a bore field of 20 T, and operating at a temperature of 20 K. | Model coil, 1m ID /2.3 m OD, 1.4 m lenght. Bore field of 20 T at 20 K operating temperature. | 2033 | 30 MCHF 37 FTEy |
| Solenoids for cooling | Split Solenoid integration demonstrator for 6D cooling cell (SOLID) | Demonstrator of HTS split solenoid performance, including integration in its support structure submitted to mechanical and thermal loads representative of a 6D cooling cell. | Target field 7 T, bore 510 mm, gap 200 mm, operating at 20 K | 2032 | 7.1 MCHF 42 FTEy |
| Solenoids for cooling | Final cooling UHF solenoid demonstrator (UHF-Demo) | Build and test a demonstrator HTS final cooling solenoid, producing 40 T in a 50 mm bore, and total length of 150 mm | 40T in a 50 mm bore, and total length of 150 mm, operated in the vicinity of liquid helium conditions, 4.5 K | 2034 | 5.6 MCHF 52 FTEy |
| RCS fast pulsed field system | RCS fast pulsed magnet string and power system (RCS-String) | Build and test a string of resistive pulsed dipoles, including powering system and capacitor-based energy storage. | Resistive dipole magnet string, +/-1.8 T field swing in a 30×100 mm aperture. Maximum ramp-rate of 3.3 kT/s, and energy recovery efficiency better than 99 % | 2032 | 6 MCHF 20 FTEy |
| LTS accelerator magnets | Wide-aperture, steady state Nb3Sn dipole for the collider (MBHY) | Demonstrate LTS dipole performance for collider arc | Prototype LTS dipole, field target of 11 T, large bore target of 160 mm, 5 m long, operating with forced-flow of helium at 4.5 K | 2036 | 11.1 MCHF 71 FTEy |
| HTS accelerator magnets | Rectangular aperture HTS dipole (MBHTS) | Demonstrate performance of rectangular aperture HTS dipole for the accelerator | Demonstrator HTS dipole, field target of 10 T, aperture of 30x100 mm, 1 m long, operating at 20 K | 2035 | 8.25 MCHF 60 FTEy |
| HTS accelerator magnets | Wide aperture HTS dipole (MBHTSY) | Demonstrate wide aperture HTS dipole for the collider arc | Demonstrator HTS dipole, field target of 14 T, large bore target of 140 mm, 1 m long, operating at 20 K | 2045 | 7.9 (15.8) MCHF 75 (126) FTEy |
| HTS accelerator magnets | Wide aperture HTS IR quadrupole (MQHTSY) | Demonstrate wide aperture HTS quadrupole for the collider IR | Demonstrator HTS quadrupole, gradient target of 300 T/m, large bore target of 140 mm, 1 m long, operating at 4.5 K | 2045 | 3.5 (8.8) MCHF 27 (60) FTEy |

A synthetic view of the lines of activity towards the eight Technology Milestones, the associated demonstrator or system test, key performance targets and parameters, the time to achievement and the required resources are reported in Tab. II. Resources are reported over the reference time span of ten years. In the case of the activities of towards TMs of longer duration, the total resources are reported in parentheses. Finally, in addition to the eight TMs, we have defined supporting material and testing activities that are specific and required to advance towards the TMs. These comprise investigating radiation resistance, mechanical properties, and cryogenic performance of HTS materials to feed magnet development and test.

A summary of the personnel resources and material cost is shown in Figs. 2 and 3, where we report both the yearly values (in FTE and MCHF/y), as well as the cumulated values (in FTE.y and MCHF). In the case of personnel, we also report the estimate of the effort that would be required to progress to the next phase of prototyping and construction, should the project proceed in this direction.

The total personnel required for the proposed magnet R&D program, over the ten-year period, is around 400 FTE y. This is rather evenly distributed, reaching 200 FTE y after five years, and a maximum just above 50 FTE in the middle of the ten-year period. To be noted that towards the second half of the ten years of R&D, proficient personnel could transfer to prototyping and construction, seamlessly absorbed in the 60 to 65 FTE that would be needed in total (including residual R&D) during this phase. Under materials we list all direct raw materials cost, consumables and services performed under contract. We include tooling in the material costs, but we exclude the cost of generic infrastructure (e.g. test infrastructure). The total material costs over ten years are estimated at 83 MCHF, with 39 MCHF expected after five years — nearly half of the total. The peak yearly expenditure, in the middle of the ten-year period, is at the level of 10 to 13 MCHF/year. Note that in this case the drop in yearly expenditure is due to the fact that a transition to the following phase of prototyping and construction is not added.

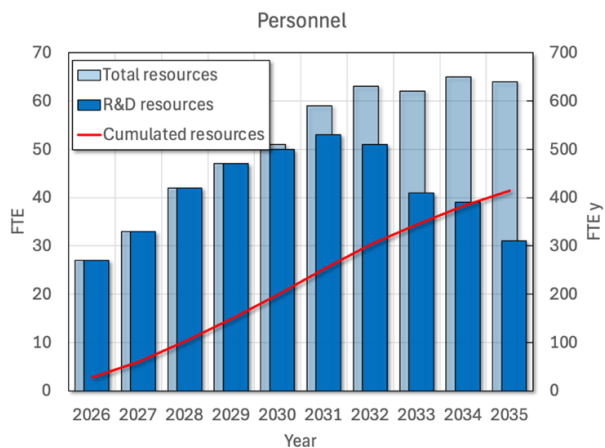 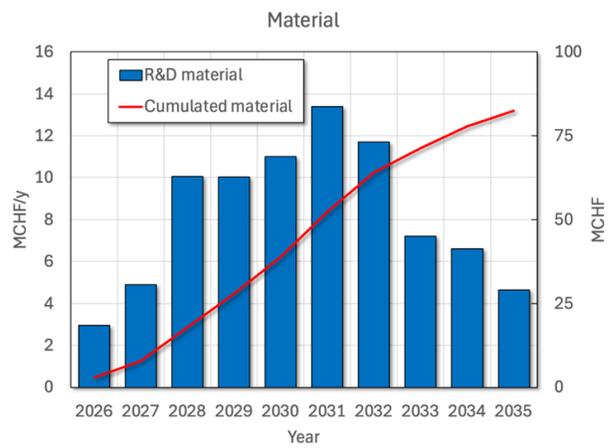

Figure 2. Time profile of the personnel resources necessary to conduct the R&D program proposed in this note, reported both as yearly values (columns, left axis), as well as cumulated over the years of activity (line, right axis). The second set of columns (shaded) reports the estimated total personnel resources that would be needed to proceed to prototyping and construction in the period after 2031.

Figure 3. Time profile of the material cost necessary to conduct the R&D program proposed in this note, reported both as yearly values (columns, left axis), as well as cumulated over the years of activity (line, right axis).

**Impact and Synergies**

The proposed R&D program is crucial to advance magnet technology to the readiness level which is required for an informed decision on the possible construction of a Muon Collider. However, this will not be the only impact of the proposed activities. We give in Fig. 4 a synoptic view of our evaluation of the impact. The representation in Fig. 4 is a purely graphical, but suggestive, demonstrating that all proposed TM's (rows) have a connection to other programmes in HEP and other fields of application (columns).

We expect that the technology developed by the R&D proposed here will impact:

- *High Energy Physics (HEP) & Future Colliders* – The technologies developed for the Muon Collider accelerator magnets are synergic with other on-going programs such as the CERN-hosted High Field Magnet development programme (HFM) [28] and the US Magnet Development Programme (US-MDP) [29]. The proposed R&D will provide technology advances that will benefit other proposed circular colliders, such as the

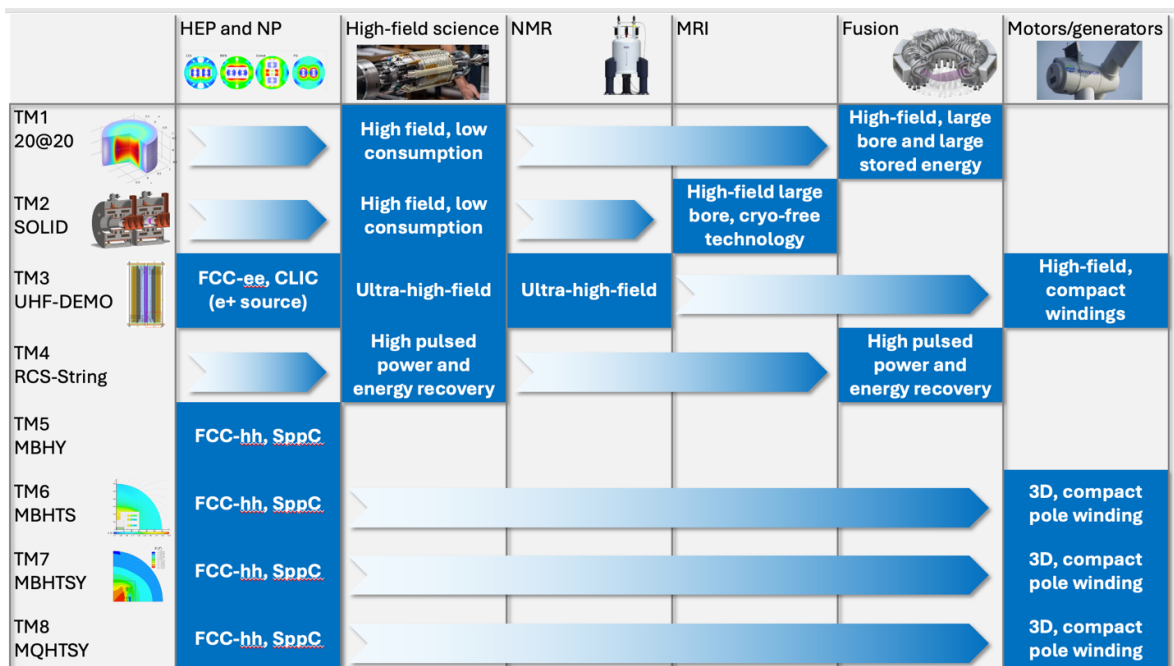

Figure 4. Schematic representation of the impact of the R&D driven by the technology milestones defined in this proposal on other programmes in HEP and other fields of application. The specific technologies of interest are indicated in the highlighted boxes.

- FCC-hh (Future Circular Collider Hadron-Hadron) and will pave the way for more efficient, high-field accelerator designs.
- *Fusion Energy* – There is a strong interest in the fusion community towards the HTS large bore solenoids required for the Muon Collider, whose application in fusion energy research will improve power plant performance and efficiency.
- *Healthcare* – MRI and NMR systems will benefit from advancements in HTS technology, towards higher fields and higher operating temperature, allowing higher resolution imaging and more energy-efficient designs.
- *Scientific Applications* – Material science and high-magnetic-field laboratories (e.g., NHMFL and EMFL) will gain access to more powerful magnet systems to extend their resolving power, revealing new states of matter.
- *Power Applications* – Compact, high-field HTS polar winding magnet technology can benefit wind energy and transportation, reducing generator and motor mass and increasing power density.

A critical aspect of the Muon Collider magnet development is ensuring energy efficiency and environmental sustainability. This is addressed through several strategies:

- *High-Temperature Superconductors (HTS) for Reduced Power Consumption* – HTS magnets operated at higher temperatures (20K) compared to traditional low-temperature superconductors, lead to significantly reduced cryogenic cooling costs and energy consumption. Furthermore, the transition from liquid helium (4.2K) to gaseous helium or cryogen-free designs reduces the risk of the helium supply chain.
- *Compact Magnet Designs for Lower Infrastructure Requirements* – Optimized magnet geometries relying on HTS non-insulated winding technology reduce the size and material needs. Smaller, more efficient systems will lead to reduced carbon footprint and infrastructure costs.
- *Energy Recovery in Rapid Cycling Magnets* – The RCS magnet system integrates capacitor-based energy storage, enabling efficient energy reuse during the rapid cycling process. Powering strategies are being devised to minimize resistive losses, reducing the overall environmental footprint.

In summary, developing the magnet technology necessary to the Muon Collider will have significant impact on the performance, energy efficiency and sustainability of research infrastructures, as well as other fields of scientific and societal application of superconducting magnets.

**Conclusion**

The development of next-generation superconducting magnet technology, with a special focus on HTS, is a critical enabler for the Muon Collider. The proposed R&D plan is designed to systematically reduce technical risks, validate key technologies, and bring magnet technology to readiness levels appropriate for large-scale deployment. While being instrumental to the Muon Collider, the broader impact extends much beyond, with the potential of significant contributions to medical imaging, fusion energy, and industrial applications. By leveraging synergies with existing HEP and industry initiatives, this R&D program is positioned to drive innovation in superconducting magnet technology and deliver a transformative leap forward in collider science.